\documentclass[10pt,twocolumn]{IEEEtran}

\usepackage{ifpdf}

\ifpdf
        \def\texver{pdftex}
\else
        \def\texver{dvips}
\fi

\usepackage{cite}
\usepackage{subfigure} 
\usepackage{stfloats}
\usepackage{amsmath}
\usepackage{amsfonts}
\usepackage{latexsym}
\usepackage[T1]{fontenc}
\usepackage{amsfonts}
\usepackage[\texver]{graphicx}
\usepackage{placeins}
\usepackage[all]{xy}
\usepackage{times}
\usepackage{fancybox}
\usepackage{latexsym}
\usepackage[pdfauthor={Yehuda Shen-bahar, Ofer Amrani},%
                        pdftitle={Iterative Multiuser Detection for Coded UWB Systems},%
                        colorlinks,citecolor=black,linkcolor=black,%
                        hypertexnames=false,%
                        \texver]{hyperref}
\usepackage{url}

\makeatletter
\def\url@leostyle{%
  \@ifundefined{selectfont}{\def\UrlFont{\sf}}{\def\UrlFont{\small\ttfamily}}}
\makeatother
\newcommand{\argmax}{\operatornamewithlimits{argmax}}

\begin{document}

\title{Graph-based Detection of  Multiuser \\ Impulse Radio Systems }
\author{Yehuda~Shen~Bahar and Ofer~Amrani\\
                School of Electrical Engineering,\\
        Dept. Elec. Eng. - systems,\\
       Tel-Aviv University, Tel-Aviv, Israel\\
       yeuda,ofera@eng.tau.ac.il, +972-3-6407766}

\maketitle

%
%
%
{\bf
{\it Abstract}  - Impulse-Radio (IR) is a wideband modulation 
technique
that 
can support multiple users by employing random Time-Hopping (TH)
combined with repeated transmissions. The latter is aimed at alleviating the impact
of collisions.
This work employs a graphical model for describing the multiuser
system
which, in turn, facilitates the inclusion of general coding
schemes. 
%
%
%
%
%
Based on factor graph representation of the system, several
iterative multiuser detectors are presented.
These detectors are applicable for any binary linear coding scheme.
The performance of the proposed multiuser detectors is evaluated via simulations revealing large gains with low complexity.
}

\noindent {\it Keywords - } 
%
\section{Introduction} 
%
%
\label{chap:intro}

%
%
%
Impulse-radio systems 
use 
short baseband pulses. In Time-Hopping Impulse Radio
(TH-IR) systems the information is encoded in the polarity of the
pulses (Pulse amplitude modulation - PAM), or in the position of
the pulses (Pulse position modulation - PPM).
%
To support multiple access, additional delay is
introduced per pulse. The added delay, which is unique for each
user, is (pseudo) random, and is assumed to be known to the receiver.
Additionally, in order to improve performance, each information
bit is typically transmitted several times.
As a result of the above measures, the probability of catastrophic
collision (multiple-user interference) is minimized.
%
%
In-depth treatment of 
Impulse Radio communications is given by~Win~and~Scholtz~\cite{1_win98impulse}.
%
While {\it impulse radio} is typically associated with Radio-Frequency (RF) communications, 
Visible Light Communications (VLC) is yet another data communication technique that can 
employ short (light) pulses for signalling, and can hence benefit from the algorithms 
and results presented herein.

TH-IR was analyzed in the past for several
channel models and interference types, see e.g.
\cite{2_narrow_int_th_ir},\cite{3_uwb_interference},
\cite{4_uwb_gaussion_noise_multipath}.
%
Scholtz \cite{5_ma_thir_scholtz} suggested employing this
communication scheme for supporting multiple access. The Multiuser
interference (MUI) was modeled as Gaussian noise and the receiver
employed a matched filter to detect a specific user
\cite{5_ma_thir_scholtz}, \cite{1_win98impulse}.
Two different receiver architectures were defined: AIRMA  (Analog
impulse radio multiple access) and DIRMA (Digital IRMA) receivers
\cite{9_airma_dirma}.
In another work \cite{6_pseudochaotic_th} the choice of a different
time  hopping sequence was discussed, a pseudo-chaotic sequence,
aimed at alleviating the MUI problem.
%
%
Deterministic sequences designed to mitigate MUI altogether were also proposed
\cite{7_all_digital_impulse_radio},\cite{8_blockspreading}.
To improve system performance and better deal with MUI in the framework of single-user detection, several authors proposed more general coding and modulation schemes,
see {\it e.g.} \cite{Multidimensional_modulation},
\cite{Coded_Modulation},\cite{Self_encoded_TH_PPM},\cite{PPAM_TH}.

%
Rather than treating the signals received from the many users as interference, one can benefit by performing {\it MultiUser Detection} (MUD), thus extracting the data associated with all users.
In general, MUD is a computationally intensive task. To alleviate this problem,
Fishler and Poor \cite{fishler-lowcomplexity} suggested using an {\it iterative}
multiuser detector for the simple repetition-based code in order to achieve good performance with low-complexity.
%
Wang {\it et. al.} \cite{wang-lowcomplexity} suggested using the
same iterative detector as \cite{fishler-lowcomplexity} with
different message passing aimed at reducing computational complexity.
Chen {\it et. al.} \cite{chen-uwbldpc} used the conventional TH-IR
scheme concatenated with low-density parity-check (LDPC) coding for
improving system performance. Their receiver employed the standard
TH-IR detector followed by an LDPC  decoder.
%
Sathish {\it et. al.} \cite{arun-codeduwb} proposed using the
standard TH-IR  scheme concatenated with convolutional coding. The
receiver employed an iterative soft-input soft-output, 3-stage
multiuser detector.
%
The aforementioned contributions relating to MUD employ concatenation of the
standard TH-IR  scheme with coding.
%
It would be interesting to introduce advanced coding \cite{Potential_UWB}
and multiuser detection techniques for further improving system performance.

%
%
%
LDPC codes were originally introduced by Gallager
\cite{1_gallager1962,2_gallager1963} in 1962, and "reintroduced" in
recent years. It was the introduction of Turbo convolutional Codes
with efficient iterative  decoding (which exhibit notably low error
probabilities at low SNR) that triggered the search for such codes -
including LDPC codes 
on this family of codes
(see e.g. \cite{3_good_error_sparse}
\cite{4_the_capacity_ldpc} \cite{5_design_capacity_ldpc}). 
In this paper we present several iterative multiuser receivers for the original TH-IR system.
Then, we study a modified system where all users employ arbitrary linear coding,
and generalize the above multiuser receiver for this setting.
%
The proposed receivers  are based on  {\it factor graph}
representation of the complete system.
Comparison with the classical repetition-based systems reveal the
significant  improvements possible with the proposed practical
approach.
%
%
%
The rest of this manuscript is organized as follows.
Section~\ref{chap:system_model} reviews the system model.
%
%
In Section~\ref{chap:mud} we present iterative multiuser detectors.
%
Finally, Section~\ref{chap:sim_result} provides simulation results
for the detectors introduced in the previous section using several
codes including LDPC codes.
Some of the results reported herein appeared in
\cite{yehuda_coded_uwb_conf}.

%
\section{System Model}
%
%
\label{chap:system_model}

%
%
%
A typical UWB TH-IR signal can be
described by the following general model:

\begin{equation}
\label{th-ir-transmit}
s_{tr}^{(k)}(t)=\sum_{j=-\infty}^{\infty}b^k_{\left\lfloor j/N_f
\right\rfloor}
        w_{tr}(t-j T_{f}-c_j^k T_c),
\end{equation}

where
$S_{tr}^{(k)}$
  is the transmitted signal of the \textit{k}'th user;
$T_f$
  is the nominal pulse repetition time;
$w_{tr}$
  is the transmitted pulse shape;
$b_{i}^{k}$
  is the $i$th symbol transmitted by user $k$;
$c_j^k$
  is the time hopping sequence used by user $k$;
$N_f$
  is the number of frames in which a symbol is transmitted;
$T_c$
  is the chip size.

A user repeats every information symbol in $N_f$ different frames,
where each frame consists of $N_c$ {\it chips}, also referred to
as {\it slots}.
The time hopping sequence employed by each user is 
a set
of values chosen randomly from $\{0, 1, \ldots, N_c-1\}$. Usually
$N_c < T_f/T_c$ 
for avoiding inter-symbol-interference (ISI).
%
In this work the information symbols are assumed to be binary
digits (bits) and the signaling is binary-phase shift keying
(BPSK).

The system consists of $K$ transmitting
users and one receiver, 
where the different users are centrally coordinated and synchronized
by the receiver (base station)
\cite{synchronizing_impulse_network},
\cite{semi_blind_synchronization},
\cite{rapid_acquisition}.
The channel over which the signal is transmitted can be frequency
selective; it is assumed that the combined response of the channel
and pulse shape is 
such that the inter-symbol interference is negligible, and the channel characteristics are slowly varying in time. The received signal is
perturbed by additive white Gaussian noise (AWGN).
%

%
%
The received signal is given by
\begin{equation}
\label{th-ir-receive}
r(t)=\sum_{k=1}^{K}A_k\sum_{j=-\infty}^{\infty}b^k_{\left\lfloor
                j/N_f \right\rfloor}
        w_{rx}(t-j T_{f}-c_j^k T_c) + n\left(t\right),
\end{equation}
where 
$K$
  denotes the number of users in the system;
$w_{rx}$
  is the received waveform associated with one transmitted pulse;
$A_k$
  is the amplitude associated with user $k$;
$n\left(t\right)$
  denotes an additive white Gaussian noise process.
%

In particular, this channel model can be encountered in scenarios where the dominant propagation path is the line-of-sight, 
and $T_c$ is chosen to satisfy $T_c>Sup\{w_{rx}\}$, where $Sup\{.\}$ is the support of the received pulse $w_{rx}$.
See also \cite{fishler-lowcomplexity} and the references therein.
%

%
\subsection{Discrete Time model}
%
%
The receiver tracks the gains associated with the different users.
The received signal \eqref{th-ir-receive} goes through a matched
filter whose output is sampled every $T_c$ seconds. Denote by
$\mathbf{r}[i]$ the vector of samples at the output of the matched filter
corresponding to the \emph{i}th information symbol. The size of
this vector is $N_fN_c$ and it can be described by the following
equation
\begin{equation}\label{eq:th-ir-receive-discr_i}
    \mathbf{r}[i] = \mathbf{S}[i]\mathbf{Ab}[i] + \mathbf{n}[i],
\end{equation}

where
  $\mathbf{S}[i]$
    is a  $[N_cN_f\times{K}]$ matrix describing the slots used by the
    different users for transmitting their \emph{i}'th information symbol. The
    matrix is defined by
    \begin{equation}\label{s-matrix}
    \mathbf{S}[i]_{lk} = \left\{ \begin{array}{ll}
        1 & \textrm{if $c^k_{(i-1)N_f+\lfloor \frac{l}{N_c} \rfloor N_c} = l-\lfloor \frac{l}{N_c} \rfloor N_c$}\\
        0 & \textrm{otherwise}.
        \end{array} \right.
    \end{equation}
    $\mathbf{A}$
    is the gain between the \emph{k}'th transmitter and the receiver, 
    $\mathbf{A}=diag(A_1,A_2, \ldots, A_k)$;
    $\mathbf{b}[i]$
    is the information vector 
    transmitted by the $K$ users at the
    \emph{i}'th 
    interval, $\mathbf{b}[i]=[b_i^1,b_i^2, \ldots, b_i^K]^T$;
    %
    $\mathbf{n}$ is a zero-mean Gaussian random vector with correlation
                  matrix $\sigma_n^2I$, where $\sigma_n^2=\frac{N_0}{2}$.

Without loss of generality we can examine the first information
symbol and therefore omit the time index $i$, consequently
\eqref{eq:th-ir-receive-discr_i} can be rewritten as
\begin{equation}\label{th-ir-receive-discr}
    \mathbf{r} = \mathbf{SAb} + \mathbf{n}.
\end{equation}

\begin{figure}[ht]
    \centering
    \includegraphics{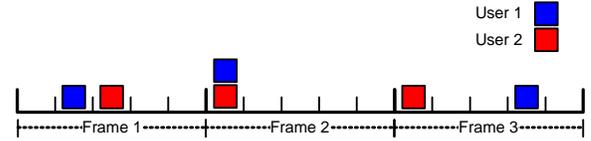}
    \caption{System illustration}
    \label{fig:SystemModel}
\end{figure}

Figure~\ref{fig:SystemModel} depicts a toy example of transmitting
one information symbol in a 2-user system with the following
parameters: $N_f=3$ frames, $N_c=5$ chips per frame. The rectangles
represent the slots used by the users.
%
%

%
\section{Iterative multiuser detection for TH-IR systems}
%
%
\label{chap:mud}

%
A major drawback of optimum Multi User Detection (MUD) is  implementation complexity.
In this section we try to alleviate this problem by presenting several iterative detectors of reduced complexity.
%
%

%
We first introduce a simple iterative detector for the original TH-IR system settings presented in Section \ref{chap:system_model}.
Then, we propose a graph-based description of the same system, and develop the corresponding MAP-based iterative detector.
%
%
%
%
%
%
%
Finally, we introduce coding into the original system
and develop the appropriate graph-based iterative multi-user detector.
%

%
%
%
%

%
%
\subsection{ID detector}
%
%
%

\label{sec:ID_decoder}
The first detector we present is 
a (very) low complexity iterative detector based on intuition, rather than mathematical arguments.
It is therefore referred to
as {\it ID} ({\it Intuition-driven}) detector.
%
It is tailored for detecting repetition-based transmissions, and may be characterized as a Gallager-type decoder.

\begin{figure}[t]
    \centering
    \includegraphics[width=200pt]{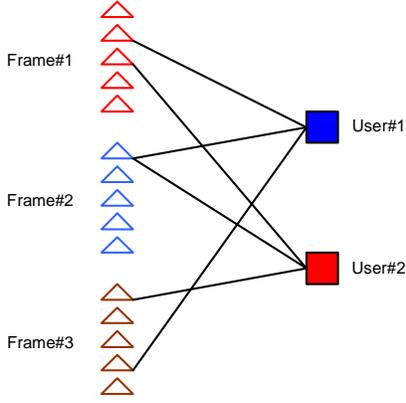}
    \vspace*{-1.5cm}
    \caption{ID decoder graph: $K=2$ users, $N_f=3$ frames, $N_c=5$ chips per frame.}
    \label{fig:simple_iterative_graph}
\end{figure}

Returning to the toy example of Figure~\ref{fig:SystemModel}, the
corresponding $\mathbf{S}$ matrix \eqref{s-matrix} is given by
\begin{equation}\label{s_for_example}
\mathbf{S}^T= \left[
    \begin{array}{ll}
    0\ 1\ 0\ 0\ 0\ \ 1\ 0\ 0\ 0\ 0\ \ 0\ 0\ 0\ 1\ 0  \\
    0\ 0\ 1\ 0\ 0\ \ 1\ 0\ 0\ 0\ 0\ \ 1\ 0\ 0\ 0\ 0  \\
    \end{array}
\right],
\end{equation}
and the 
graph we associate with the ID detector is given in Figure~\ref{fig:simple_iterative_graph}.

The detector
consists of two stages performed in an iterative manner.
In the first stage,
the estimated value of a specific pulse is calculated given the
received signal and the information transmitted by other colliding
users.
Associated with the first stage of the decoding is the left-hand side of the
graph consisting of at most $N_f N_c$ nodes,
one for each of the outputs $r$ of the matched filter. We shall hence refer to these nodes as \emph{input nodes}.
In the second stage,
the estimated value of a bit associated with a specific user
is calculated given the information from the first stage.
The second stage uses the fact that all the pulses of the
\emph{k}th user correspond to the same bit.
The right hand side of the graph is associated with the second stage of the decoding. It consists of $K$ nodes, referred to as \emph{check nodes}, one for each of the $K$ users.

The connection between the two sets of nodes is determined by the matrix $\mathbf{S}$ -
\emph{input node} $i$ is connected to \emph{check node} $j$, if
$S_{ij}=1$.
%
The messages passed between the nodes
will are "hard" (binary valued), rather than "soft" (e.g. LLR).

\begin{figure}[t]
\centering
\subfigure[ID decoder - input node]{
            \includegraphics[width=100pt]{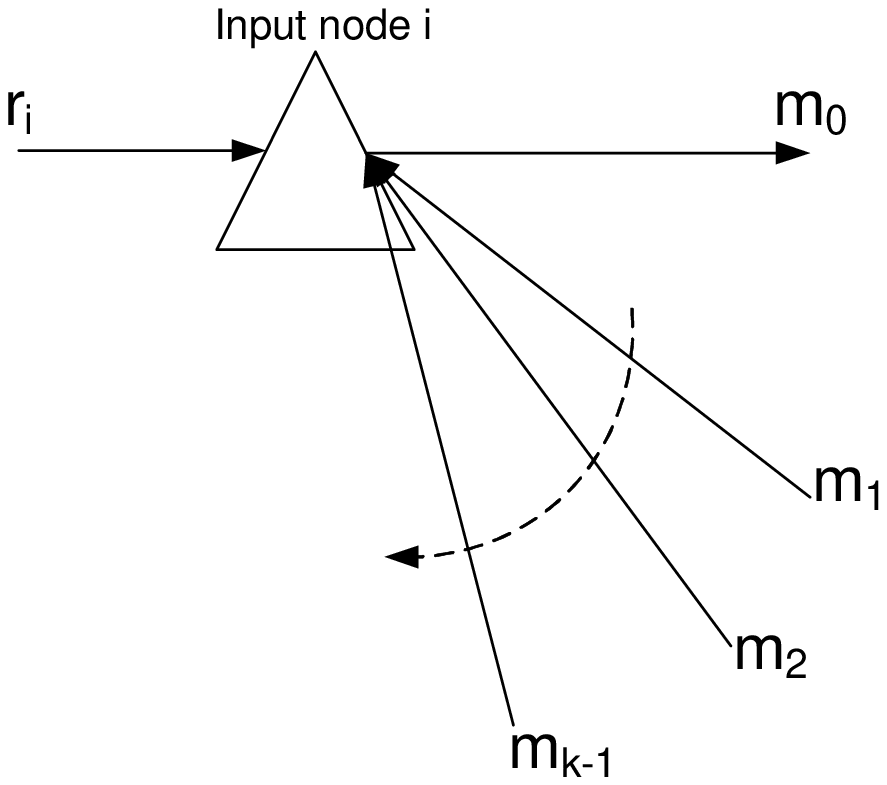}
            \hspace*{1.0cm}
            \label{fig:dummy_input_node}
            }
\subfigure[ID decoder - check node]{
            \includegraphics[width=79pt]{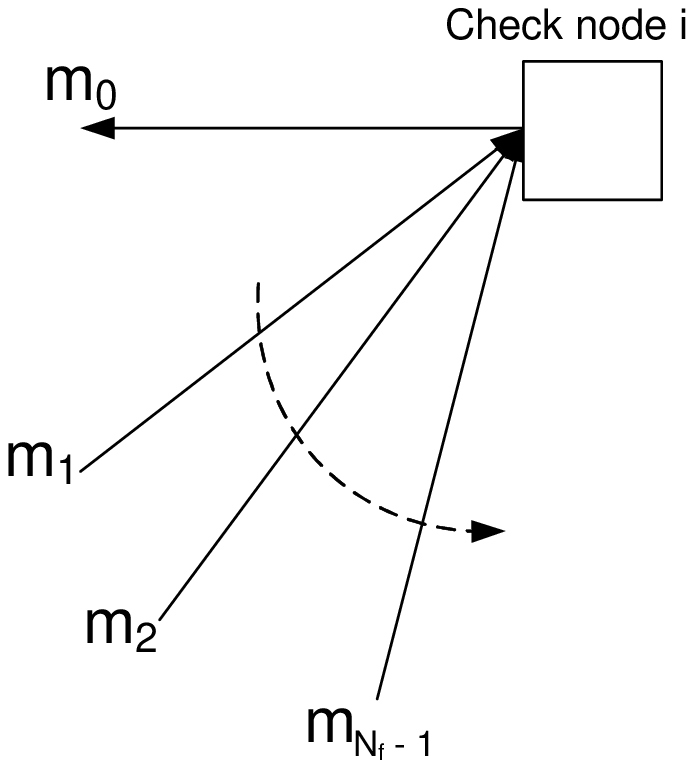}
            \hspace*{-0.5cm}
            \label{fig:dummy_check_node}
            }
\label{fig:id_decoders_nodes}
\caption{ID decoder nodes.
                }
\end{figure}

%
We shall now provide a quantitative description of the ID detector 
beginning with the input nodes, one of which~is~depicted~in~Figure~\ref{fig:dummy_input_node}.
%
%
Without loss of generality consider an output message
passed via the first edge. Let $m_0$ denote the
output message, while $m_j, j\neq 0$, denote all the input
messages, and $r_i$ is the output of the matched filter connected
to input node $i$.
Note that an input node is connected to $k$
check nodes, $0 \leq k \leq K$ being the number of users choosing to
transmit in chip position \emph{i}
\begin{equation}\label{j_value}
    k=\sum_{l=1}^{K}S_{i,l}.
\end{equation}
The output message passed to the check node is defined as
\begin{equation}\label{dummy_input_node_message}
    m_0 = r_i-\sum_{l=1}^{k-1}A_{i,l}*m_l,
\end{equation}
where $A_{i,l}$ is the amplitude 
of user \emph{l} as seen by input node $i$.


With the aid of Figure~\ref{fig:dummy_check_node}, we proceed to describe the check nodes.
Any check node is connected to exactly $N_f$ input nodes (due to the repetition nature of the transmission).
The message computed by a check node (to be passed back to an input node) is defined as
\begin{equation}\label{dummy_check_node_message}
    m_0 = sgn( \sum_{l=1}^{N_f-1}m_l ).
\end{equation}
Note that for the first iteration, the output from all check nodes is initialized to zero.
Finally, after the last iteration, the output of the
detector is 
taken from the check nodes, yet it includes all the inputs as follows
\begin{equation}\label{dummy_check_node_last}
    \hat{b_k} = sgn( \sum_{l=0}^{N_f-1}m_l ).
\end{equation}

%
\subsection{3-stage Factor-Graph (FG3) detector}
%
%
%
\label{sec:FG3_Decoder}

In this section we present an iterative (soft) detector based on a
3-stage factor-graph description of the system.
While the proposed detector is aimed at decoding the repetition-based transmission,
it is of great interest as it lays the ground for
introducing arbitrary linear coding into the system.
%

Henceforth, 
we let $\mathbf{y}$ represent the received vector $\mathbf{r}$, i.e. $\mathbf{y}=\mathbf{r}$.
The output of the MAP decoder (for the $k$th bit) is given by
\begin{equation}\label{MAP_decoder}
\begin{split}
    \hat{b}^k &= \argmax_{b^k=\pm1}\{p(b^k | \mathbf{y})\} \\
          &= \argmax_{b^k=\pm1} \left\{ \sum_{-b^k}p(\mathbf{y} | \mathbf{b})
                                                    *\frac{p(\mathbf{b})}{p(\mathbf{y})} \right\},
\end{split}
\end{equation}
where $\sum_{-b^k}$ denotes summing over all values of the vector $\mathbf{b}$
excluding those containing $-b^k$.
Since all input vectors, $\{\mathbf{b}\}$ are equiprobable,
%
\begin{equation}\label{MAP_decoder_1}
\begin{split}
    \hat{b}^k &= \argmax_{b^k=\pm1}\left\{\sum_{-b^k}p(\mathbf{y} | \mathbf{b}) \right\}\\
          &= \argmax_{b^k=\pm1}\left\{\sum_{-b^k}p(\mathbf{y} | b^1,\ldots,b^K)\right\}. \\
\end{split}
\end{equation}

Using the following definition
\begin{equation}\label{E_define}
    \mathbb{E}(x_1,\ldots,x_i) =
            \left\{ \begin{array}{ll}
            1 & if x_1=\ldots=x_i \\
            0 & otherwise
            \end{array}, \right.
\end{equation}
one can write
\begin{equation}\label{MAP_decoder_2}
\begin{split}
    \hat{b}^k_j = \argmax_{b^k_j=\pm1} \{ \sum_{-b^k} &p(\mathbf{y} |
    b_0^1,\ldots,b_{N_f-1}^1,\ldots,b_K^1,\ldots,b_{N_f-1}^K)
                                       \\&
     \cdot \prod_{k=1}^{K}\mathbb{E}(b_0^k,\ldots,b_{N_f-1}^k)
     \}.
\end{split}
\end{equation}
Note that the subscript $j$, denoting the frame index, has been
added though one expects all $N_f$  estimations $\hat{b}_j^k, j\in\{0,1,...N_f-1\}$ to
produce the same value.

Since the channel is memoryless
\begin{equation}\label{memoryless_channel}
    p(\mathbf{y} | b_0^1,\ldots,b_{N_f-1}^1,\ldots,b_K^1,\ldots,b_{N_f-1}^K)
                    = \prod_{i=1}^{N_fN_c-1}p(y_i | b_{yi}),
\end{equation}
where $b_{yi}$ represents the transmitted bits from all users
employing 
chip slot $i$.
Finally, we have
\begin{equation}\label{MAP_decoder_3}
    \hat{b}^k_j = \argmax_{b^k_j=\pm1}\left\{\sum_{-b^k} \prod_{i=1}^{N_fN_c-1}p(y_i|b_{yi})
    \prod_{k=1}^{K}\mathbb{E}(b_0^k,\ldots,b_{N_f-1}^k)\right\}.
\end{equation}

\begin{figure}[t]
\centering
\subfigure[FG3 based detector]{
            \includegraphics[width=100pt]{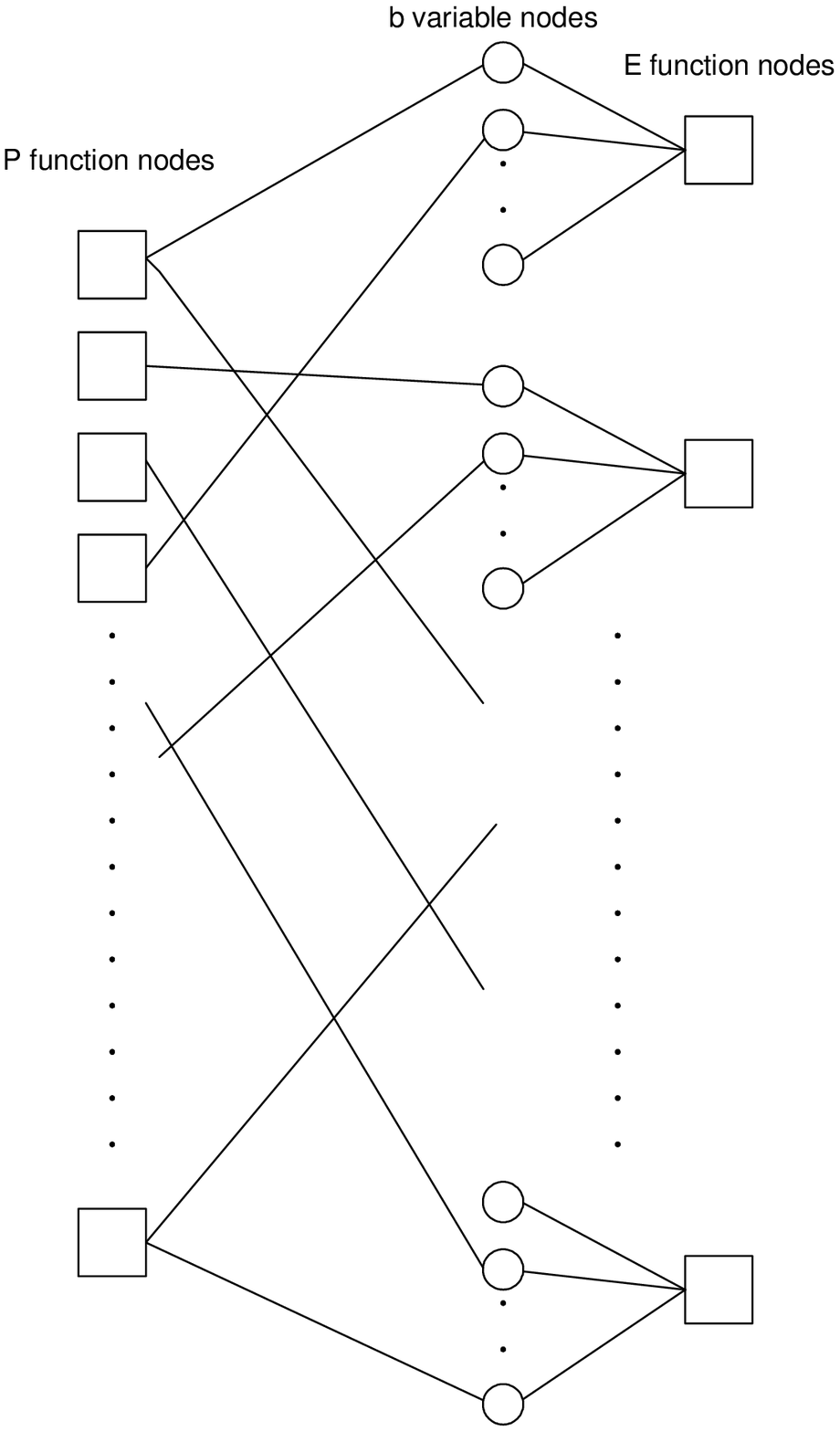}
            \label{fig:factor_graph_full}
            }
\subfigure[CFG3 based detector]{
            \includegraphics[width=100pt]{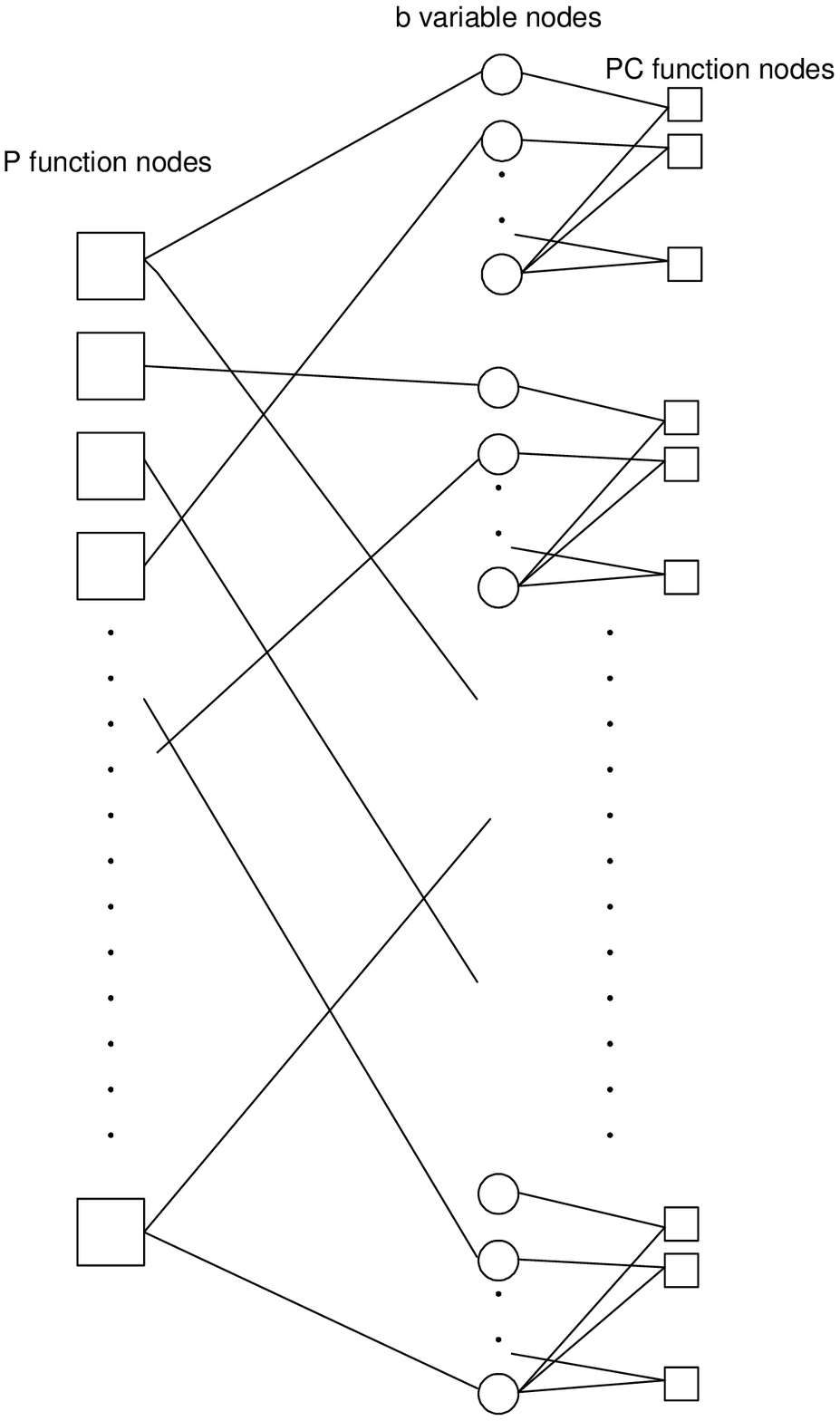}
            \label{fig:factor_graph_full_h_repetion}
            }
\label{fig:input_dist_per_users}
\caption[Factor Graph representation for Multiuser Detection ]
                {Factor graph representation used for multiuser detection - Repetition code example.
                }
\end{figure}

This, so-called, "sum-of-products" function can be calculated
using a massage passing algorithm operating iteratively
on a bipartite graph, termed {\it factor graph} \cite{kschischang01factor},\cite{factor-intro}.
%
A factor graph 
typically consists of two types of nodes - {\it function nodes} and {\it variable nodes}.
In Figure~\ref{fig:factor_graph_full} we depict a graph
corresponding to Equation \eqref{MAP_decoder_3}.

%
%
%
%
The graph consists of three types of nodes.
The squares represent function nodes, which are divided into two types.
The first type of function nodes are associated 
with the function $p(y_i|b_{yi})$; there are exactly $N_fN_c$ such nodes.
We shall denote these 
as \emph{P function nodes}.
The second type of function nodes are associated with 
$\mathbb{E}(b_0^k,\ldots,b_{N_f-1}^k)$;
there are exactly $K$ such nodes.
Denote this type of nodes as \emph{E function nodes}.
Nodes indicated by circles represent the variable nodes for
$b_i^k$; there are exactly $KN_f$ such nodes.
Denote these nodes as \emph{b variable nodes}.
Note that the edges connecting the \emph{P function nodes} with
the \emph{b variable nodes} are completely defined by the matrix $S$.
%
%
Finally, as will be shown later on, by omitting the \emph{b variable nodes}, 
the graph 
reduces to the one shown 
for the ID detector.

The messages passed by the algorithm 
shall be denoted by $\mu(x)$, $x=\pm1$.
%
$\mu_{p\rightarrow v}$ represents a message passed from a \emph{P function node} to a \emph{b variable node},
$\mu_{v\rightarrow E}$ represents a message from \emph{b variable node} to \emph{E function node},
while $\mu_{E\rightarrow v}$ is a message from \emph{E function node} to \emph{b variable node}.
Next, we provide explicit description of the messages associated
with the different types of nodes.

\vspace*{0.7cm}
%
\subsubsection{\emph{\it P function nodes}} 
%
%

The $\mu_{p \rightarrow v}$ messages, calculated at the \emph{P function nodes},
are 
\begin{equation}\label{p_function_message}
\begin{split}
    \mu_{p \rightarrow v}(+1)  &= \sum_{x_1,\ldots,x_J}p(y|+1,x_1,\ldots,x_{J})\prod_{j=1}^{J}\mu_j(x_j); \\
    \mu_{p \rightarrow v}(-1) &= \sum_{x_1,\ldots,x_J}p(y|-1,x_1,\ldots,x_{J})\prod_{j=1}^{J}\mu_j(x_j),
\end{split}
\end{equation}
where $x_i$ is a binary value conveyed by an edge connected to an adjacent 
variable node. 
W.l.o.g. $\mu(\pm1)$ denotes the output message
passed to the variable node of index $0$, 
while
$\mu_j(x_j)$ denotes an input message originating at an adjacent \emph{b variable node}. (There are $J$ input edges, with $j$ being the edge index.)
We shall henceforth omit the cumbersome directive pointers, $p \rightarrow v$ and $v \rightarrow p$, as it will always be easy to realize the correct flow. 

Rather than
passing two messages, one may use 
a single
message in the form of the log likelihood ratio (LLR).
Let $r$ 
be defined as follows
\begin{equation}\label{r_l_define}
\begin{split}
    r   & \equiv \frac{\mu(+1)}{\mu(-1)} \\
%
%
          &= \frac{ \sum_{x_1,\ldots,x_J}p(y|1,x_1,\ldots,x_{J})
                                        \prod_{j=1}^{J}\mu_j(x_j)}
                { \sum_{x_1,\ldots,x_J}p(y|-1,x_1,\ldots,x_{J})
                            \prod_{j=1}^{J}\mu_j(x_j)} \\
        &= \frac{ \sum_{x_1,\ldots,x_J}p(y|1,x_1,\ldots,x_{J})
                                    \prod_{j=1}^{J}\mu_j(x_j)\frac{1}{\mu_j(-1)}}
                { \sum_{x_1,\ldots,x_J}p(y|-1,x_1,\ldots,x_{J})
                        \prod_{j=1}^{J}\mu_j(x_j)\frac{1}{\mu_j(-1)}}.
\end{split}
\end{equation}

Note that
\begin{equation}\label{r_mu_relation}
    \frac{\mu_j(x_j)}{\mu_j(-1)} = r_j^{\frac{x_j+1}{2}}.
\end{equation}
%
%
%
%

Finally, using \eqref{r_mu_relation} the LLR, $l=log(r)$, is given by

\begin{equation}\label{p_function_message_l}
    l   = \log{
            \frac{ \sum_{x_1,\ldots,x_J}p(y|1,x_1,\ldots,x_{J})
                                \prod_{j=1}^{J}{e^{l_j}}^{\frac{x_j+1}{2}}}
                 { \sum_{x_1,\ldots,x_J}p(y|-1,x_1,\ldots,x_{J})
                            \prod_{j=1}^{J}{e^{l_j}}^{\frac{x_j+1}{2}}}}.
\end{equation}

\vspace*{0.7cm}
%
%
\subsubsection{\emph{\it b variable nodes}} 
%
%
It is easy to see that each of the \emph{b variable nodes} is
connected to exactly one \emph{P function node} on the left and
one \emph{E function node} on the right. In this case the output
messages are the same as the input messages.

\vspace*{0.7cm}
%
%
\subsubsection{\emph{\it E function nodes}} 
%
%
For this type of nodes we have 
%
%
%
%
\begin{equation}\label{e_function_message_1}
\begin{split}
    r   &= \frac{ \sum_{x_1,\ldots,x_{N_f-1}}\mathbb{E}(1,x_1,\ldots,x_{N_f-1})
                                            \prod_{j=1}^{N_f-1}\mu_j(x_j)}
                    { \sum_{x_1,\ldots,x_{N_f-1}}\mathbb{E}(-1,x_1,\ldots,x_{N_f-1})
                                    \prod_{j=1}^{N_f-1}\mu_j(x_j)} \\
        &= \frac{ \sum_{x_1,\ldots,x_{N_f-1}}\mathbb{E}(1,x_1,\ldots,x_{N_f-1})
                                        \prod_{j=i}^{N_f-1}r_j^{\frac{x_j+1}{2}}}
                    { \sum_{x_1,\ldots,x_{N_f-1}}\mathbb{E}(-1,x_1,\ldots,x_{N_f-1})
                                    \prod_{j=i}^{N_f-1}r_j^{\frac{x_j+1}{2}}.}
\end{split}
\end{equation}
Recalling the definition of
$\mathbb{E}$, \eqref{E_define}, it is easily verified that
\begin{equation}\label{e_function_message_r}
\begin{split}
    r   &= \frac{ \prod_{j=1}^{N_f-1}r_j^{\frac{1+1}{2}}}
                      { \prod_{j=1}^{N_f-1}r_j^{\frac{-1+1}{2}}} \\
        &= \prod_{j=1}^{N_f-1}r_j,
\end{split}
\end{equation}
and the LLR is simply
\begin{equation}\label{e_function_message_l}
    l   = log(r) = \sum_{j=1}^{N_f-1}l_j.
\end{equation}
After the last iteration is performed, the output of the detector is
taken from the \emph{E function node}:
\begin{equation}\label{b_k_output_fg3_rep}
    \hat{b}^k = sgn(\sum_{j=1}^{N_f}l_j).
\end{equation}



We described a multiuser detector based on 
factor graph representation of the system.
Recall that Fishler and Poor (FP) \cite{fishler-lowcomplexity} presented an iterative multiuser detector
 for the same system that follows the turbo principle.
%
%
Interestingly, the FG3 detector turns out to be the same as the FP detector
although a different model is employed for describing the system.
This assertion may not be obvious by simply comparing at the massage
passing equations. However, by using simple mathematical
manipulations one can move from the set of equations presented
herein to those used by FP.
It can be shown that the FP and FG3 detectors are not MAP-achieving
since the associated graphs are not cyclic-free.
%

%
\subsection{CFG3 detector}
%
%
\label{sec:CFG3_decoder}

The FG3 proposed model and detection technique shall now 
serve as the basis for the introduction of a coded system.
The receiver to be developed in this section
is aimed at providing a solution for a system employing
arbitrary linear coding. It
shall hence be denoted as coded-FG3 (CFG3). 
%
%

A linear code is typically defined by its parameters $[n,k,d]$,
where $n$ is the code length, $k$ is the dimension of the code (not to be confused with the number of users),
and $d$ denotes the minimum Hamming distance of the code.
Hence, in a coded TH-IR system, the number of frames $N_f$ will satisfy $N_f=n$,
and the system rate is $\frac{k}{n}$.

Referring to Equation \eqref{MAP_decoder_3}, the function $\mathbb{E}(\cdot)$
represents the fact that the bits associated with a
specific user must all be the same.
%
When a code $[n,k,d]$ is used, we shall replace $\mathbb{E}(\cdot)$
with a new function, $\mathbb{PC}$, representing a Parity Check
equation.

The parity check matrix $\mathbf{H}$ is an $(n-k)\times n$ binary
matrix
\begin{equation}\label{pc_function_def}
    \mathbb{PC} = \prod_{i=1}^{n-k}PC_i,
\end{equation}
where $PC_i$ is defined as follows
\begin{equation}\label{pci_function_def}
    \mathbb{PC}_i = \left\{ \begin{array}{ll}
            1   &   \textrm{if ($\sum_{j,h_{i,j}=1}b_j)=1$ where the sum is over GF2} \\
            0   & \textrm{otherwise}
        \end{array}. \right.
\end{equation}

As an example, Figure~\ref{fig:factor_graph_full_h_repetion} depicts
a factor graph of a linear code when the latter is a repetition code
whose
parity check matrix $\mathbf{H}$ is given by
\begin{displaymath}
\mathbf{H} = \left(
\begin{array}{ccccc}
1 & 0 & 0 & \ldots & 1 \\
0 & 1 & 0 & \ldots & 1 \\
\vdots & \vdots & \ddots \\
0 & 0 & \ldots & 1 & 1  \\
\end{array}
\right).
\end{displaymath}
%

Although this graph is different from Figure~\ref{fig:factor_graph_full}, which means that two different detectors are to be used, both graphs target the same system and code.

The \emph{b variable nodes} have one input (form the left), but
typically several outputs (connected to PC function nodes).
The messages to be 
calculated and passed to the output 
are
\begin{equation}\label{b_varible_message}
\begin{split}
    \mu(1)  &= \prod_{i}\mu_i(1) \\
    \mu(-1) &= \prod_{i}\mu_i(-1),
\end{split}
\end{equation}
and hence
\begin{equation}\label{b_varible_message_rl}
\begin{split}
    r = \frac{\mu(1)}{\mu(-1)} = \prod_{i}r_i \\
    l = log(r) = \sum_{i}l_i.
\end{split}
\end{equation}


%
The messages to be calculated at the \emph{PC function
nodes} 
are associated with single parity check equations.
 %
%
These are the same 
messages used 
in LDPC decoding:
\begin{equation}\label{pc_function_node_l}
    l = 2\tanh^{-1}(\prod\tanh\frac{l_j}{2}).
\end{equation}
After iterating all messages through the graph, the final
marginalization is performed at the \emph{b variable nodes}
\begin{equation}\label{b_k_output_fg3_rep_final}
    \hat{b}^k = sgn(\sum_{j}l_j).
\end{equation}
%

It can be shown that by appropriately choosing the code, the
associated graph can be made cyclic-free, and hence CFG3 is
MAP-achieving.

\section{Simulation Results}
%
%
\label{chap:sim_result}

%
%
\subsection{Conventional repetition-based system}
%
%
%
%
We first study the original system 
employing a repeated transmission scheme. We shall consider two
different detectors: the FG3 detector and the CFG3
detector. In all presented simulations the number of slots per frame
is $N_c = 20$, while the number of users varies, $K=\{3,10,30\}$.

\begin{figure}[t]
    \centering
        \includegraphics[width=\columnwidth]
            {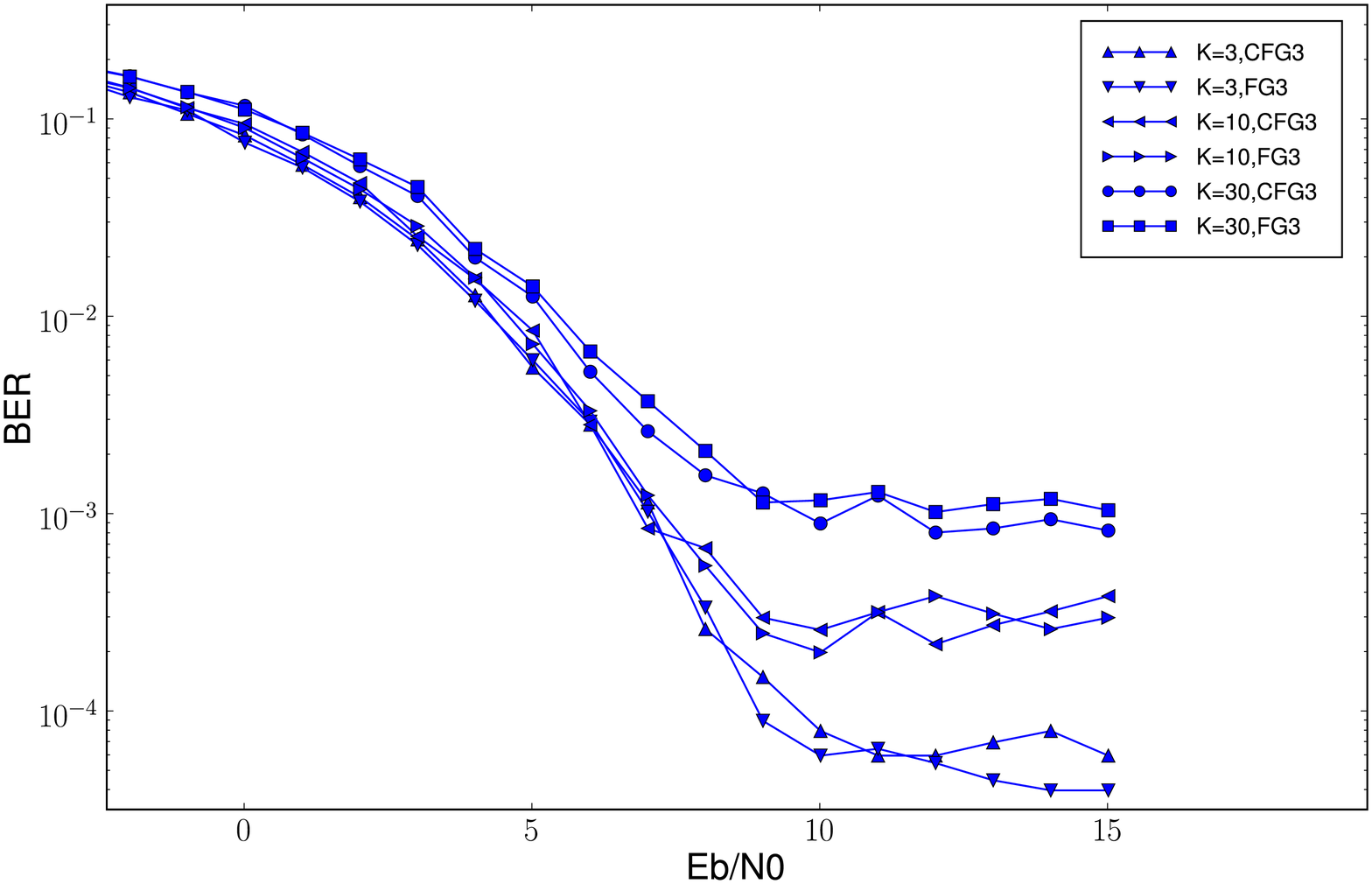}
    \vspace*{-0.8cm}
    \caption{Repetition-based system, FG3 and CFG3 detectors, $Nc=20$}
    \label{fig:eb_n0_repetions_compare}
\end{figure}
Figures \ref{fig:eb_n0_repetions_compare} hold the
simulation results.
The performance of the FG3 and the CFG3 detectors are practically the
same.  
Therefore, whenever using repetition-based coding,
we shall only consider the FG3 detectors.

%
%
\subsection{Repetition vs. LDPC-based systems}
%
%
%
We next compare the performance of a repeated transmission system
(using ID and FG3 detectors) with an LDPC-based coded system (using
CFG3 detector).
%
%
The parameters of the LDPC code chosen are $(n=120,k=56,R=0.4667)$ -
Mackay code 120.64.3.109 \cite{MackayCodes}. It has been chosen
because of its relatively short length\footnote{Recall that the
number of frames, $N_f$, in our system model is actually the
code length. Since the total number of time slots 
in the system is $N_fN_c$, reasonable simulation times are when
using shorter codes.};
no attempt has been made to optimize it for our system.
%
%
Repetition code of rate $\frac{1}{3} \rightarrow N_f=3$ have been
chosen for comparison.

\begin{figure}[t]
    \centering
        \includegraphics[width=\columnwidth]
            {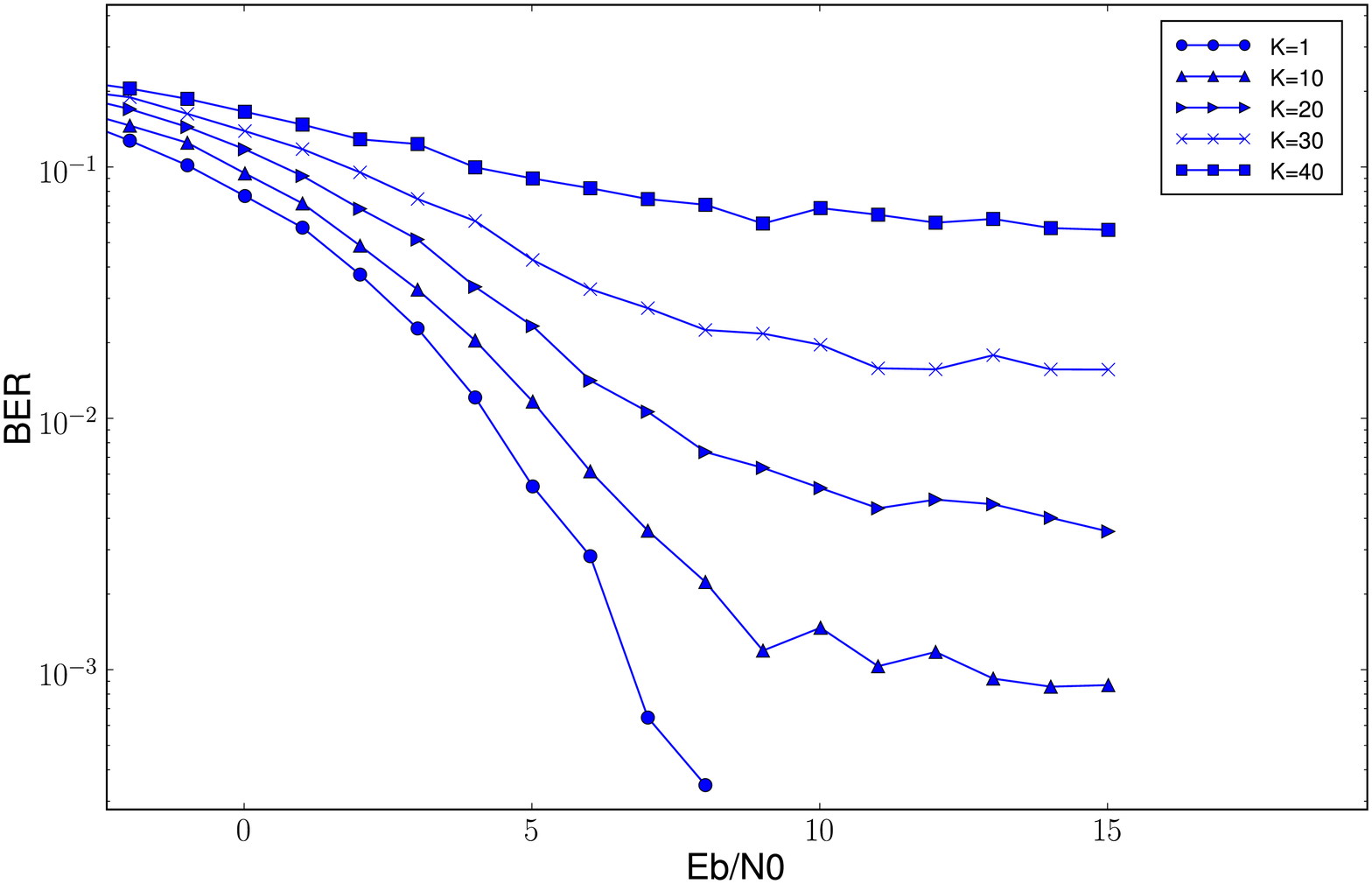}
    \vspace*{-0.8cm}
    \caption{ID detector, Repetition, $N_f=3$, $N_c=20$}
    \label{fig:eb_n0_dummy_repetions}
\end{figure}

\begin{figure}[t]
    \centering
        \includegraphics[width=\columnwidth]
            {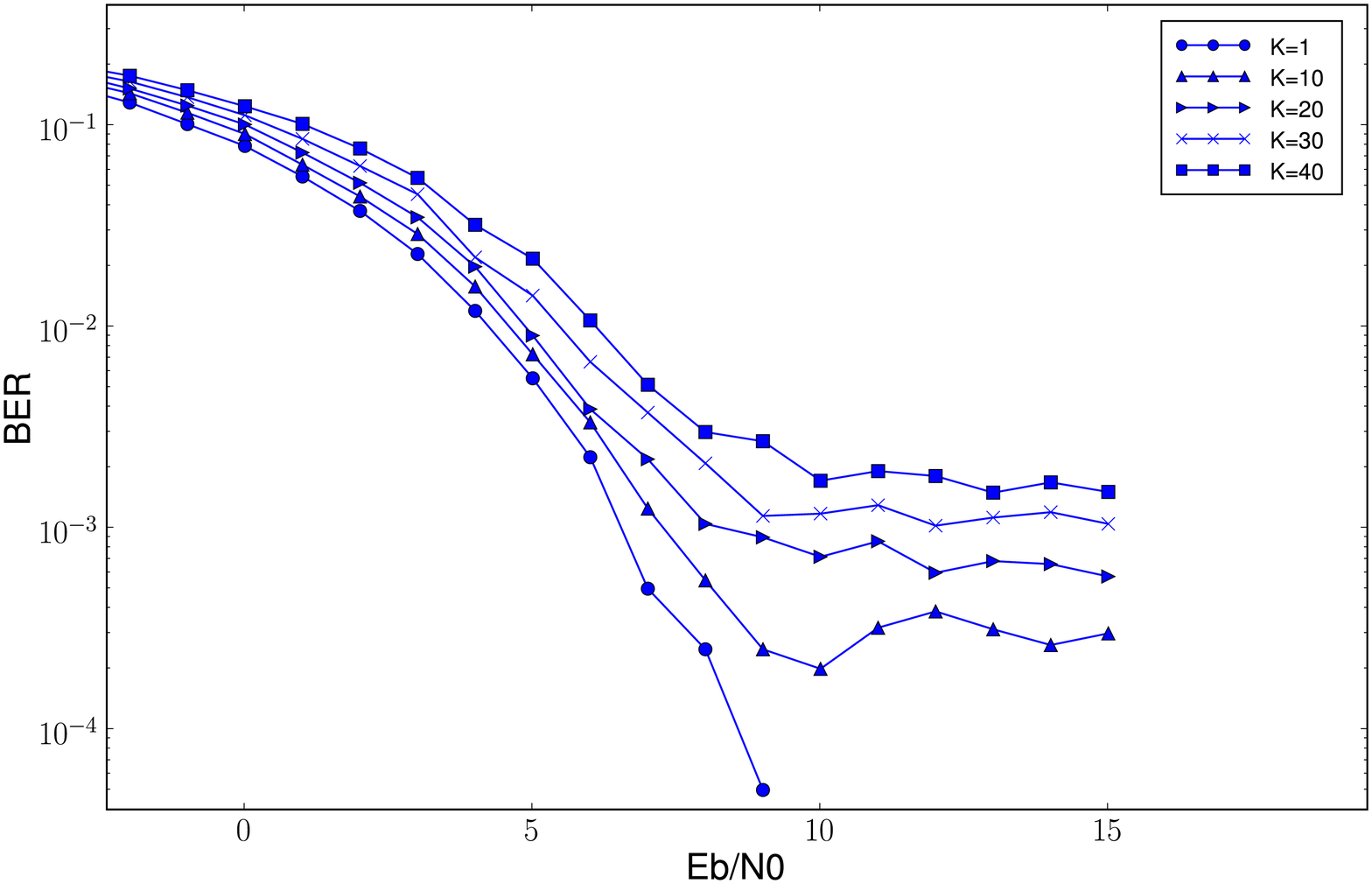}
    \vspace*{-0.8cm}
    \caption{FG3 detector, Repetition,  $N_f=3$, $N_c=20$}
    \label{fig:eb_n0_fg3_repetions}
\end{figure}

\begin{figure}[t]
    \centering
        \includegraphics[width=\columnwidth]
            {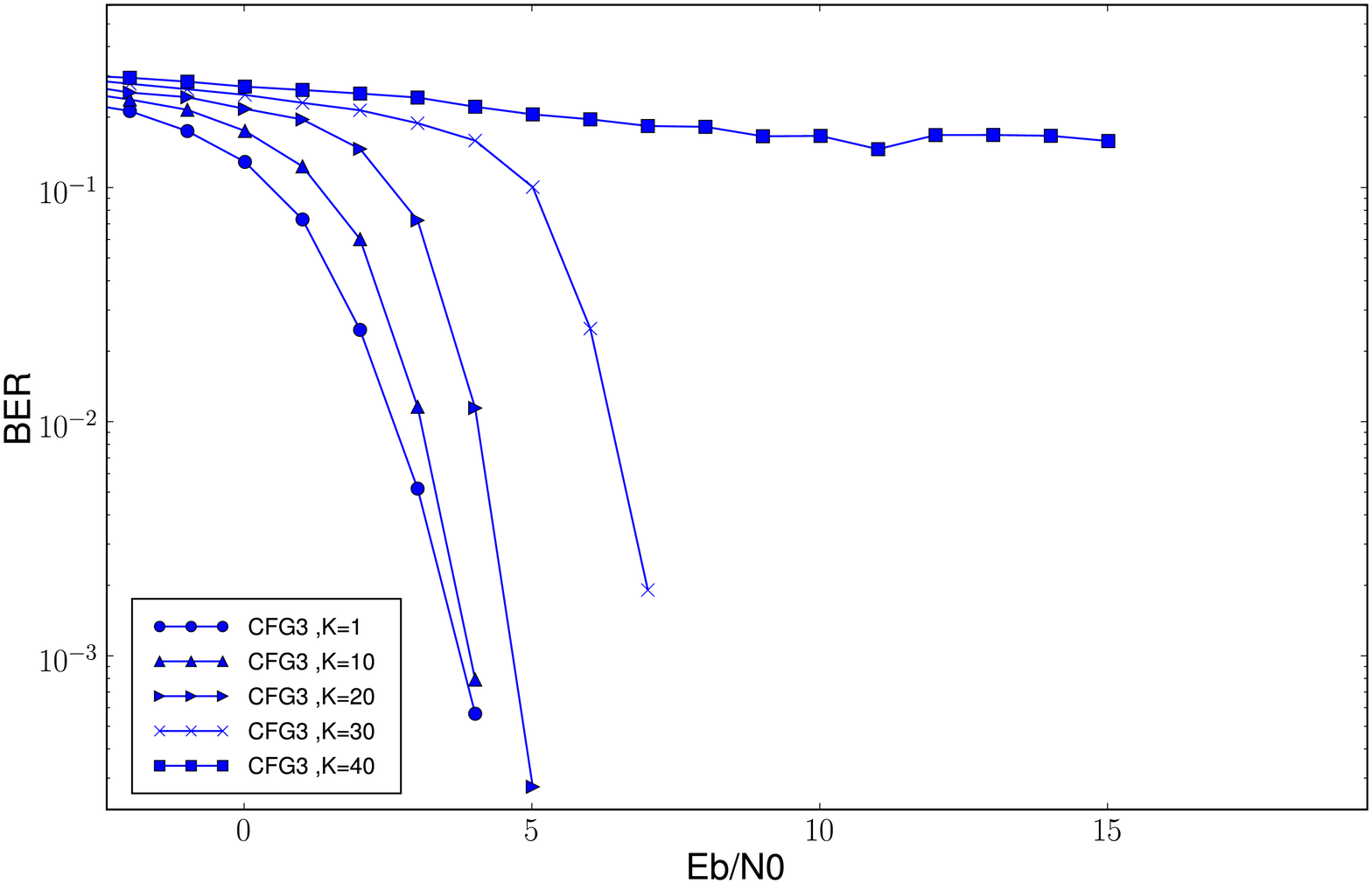}
    \vspace*{-0.8cm}
    \caption{CFG3 detector, LDPC, ($n=120$,$k=56$,$R=0.4667$), $N_c=20$}
    \label{fig:eb_n0_fg3gh_ldpc}
\end{figure}

Figures~\ref{fig:eb_n0_dummy_repetions} and
\ref{fig:eb_n0_fg3_repetions} pertain to 
the repetition code with ID and FG3 detectors, respectively, while
Figure~\ref{fig:eb_n0_fg3gh_ldpc} pertains to the LDPC code-based
system employing the CFG3 detector.
 %
The ID and FG3 detectors exhibit an error floor which
is irreducible due to the fact that system
performance is limited by Multi User Interference (MUI). To support
this assertion note how the error floor increases with the number of
users.
%
The FG3 detector exhibits better error floor performance than the ID
detector, as expected.

Clearly, the LDPC-based system with the CFG3 detector behaves
differently.
First, the BER curve is much sharper as might be expected of a 
coded system. 
Second, as demonstrated in Figure \ref{fig:eb_n0_fg3gh_ldpc}, when
the number of users increases, the $E_b/N_0$ threshold-point also
increases. Still, once the threshold is passed, the slopes
associated with all cases are similar.
%
Increasing the number of users amounts to adding more noise, which
leads us to the next observation:
another threshold that can be clearly identified corresponds to the
number of users - beyond a certain number of users, the system
collapses.
%

In general, we argue that the LDPC-based system handles MUI much
better than the alternative approaches mainly because it employs a
large number of frames ($N_f$). Consequently, "catastrophic"
collisions among different users are much less common.
Complexity-wise we argue as follows.
ID performs only simple addition operations.
Calculation of the message at the input node \eqref{dummy_input_node_message} requires $O(K`)$ additions,
%
where $K`$ is the number of colliding users. 
Check node calculation \eqref{dummy_check_node_message} requires 
$O(N_f)$ additions.
Recall that the graphs associated with the ID and FG3 systems are the same.
The two detectors differ only in the 
calculations carried out at the input nodes. 
Calculating \eqref{p_function_message_l}, instead of \eqref{dummy_input_node_message}, 
increases complexity by $O(2^{K`})$ \cite{fishler-lowcomplexity}.
Moving from FG3 to CFG3,
as the input nodes remain the same,
the added complexity is that of decoding.
In our example, LDPC coding has been employed, and therefore the additional complexity 
is given by $C_{LDPC}\cdot K$, where $C_{LDPC}$ is the  complexity required for decoding the LDPC code used.

Compared to ID, FG3 better treats MUI for the price of increased complexity at the input side (left-hand side of the graph).
CFG3 add coding gain for the price of increased complexity at the coding side (right-hand side of the graph). 
In conclusion, among the three described schemes (for the same
$E_b/N_0$), the LDPC-based system provides the best performance both
in terms of BER and overall system throughput.
%
%

%
%
\subsection{Performance dependency on the number of users}
%
%
%
In a multi-user environment users often join and leave the system.
Analyzing the performance of the system for different numbers of
users is therefore of interest.

\begin{figure}[t]
    \centering
        \includegraphics[width=\columnwidth]
            {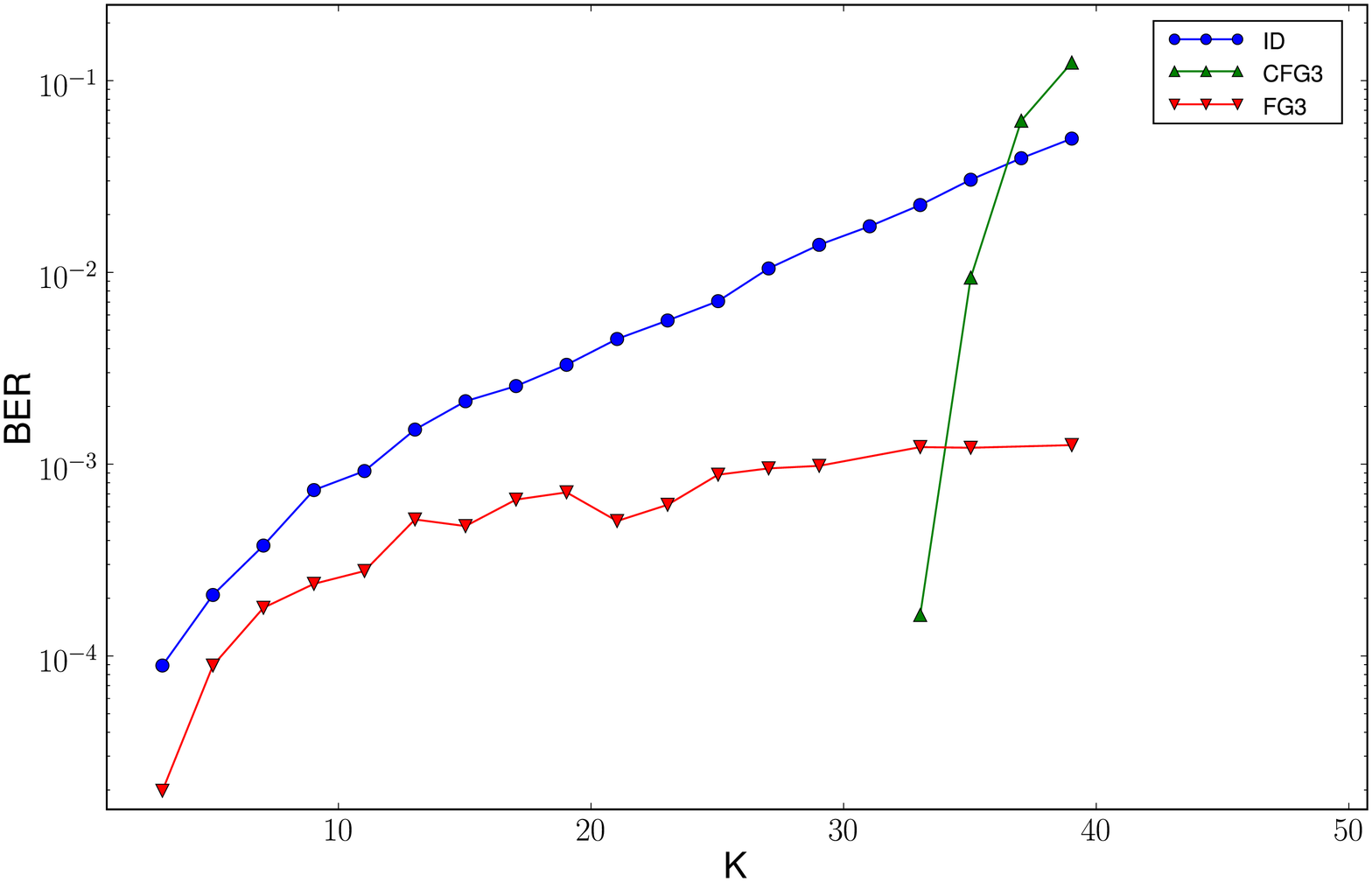}
    \vspace*{-0.8cm}
    \caption{Multi-User Performance, $E_b/N_0=20dB$}
    \label{fig:ber_users_all_snr20}
\end{figure}

\begin{figure}[t]
    \centering
        \includegraphics[width=\columnwidth]
            {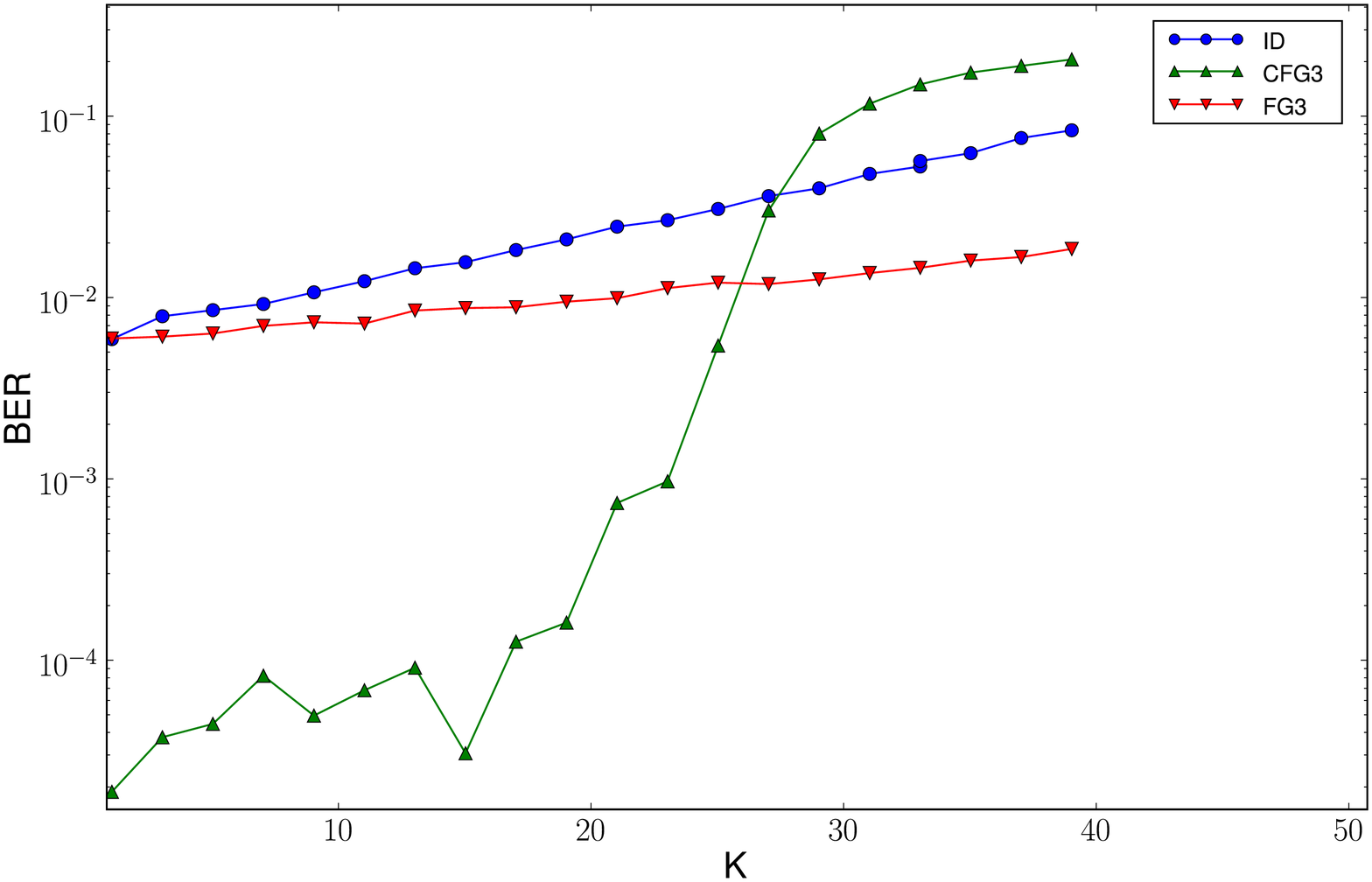}
    \vspace*{-0.8cm}
    \caption{Multi-User Performance, $E_b/N_0=5dB$}
    \label{fig:ber_users_all_snr5}
\end{figure}

A system with $N_c=20$ slots per frame is considered:
ID and FG3 detectors are employed for a repetition-based
transmission scheme with $N_f=3$ $(rate=1/3)$; CFG3 detector with
LDPC code $(n=120$,$k=56$,$R=0.4667)$ is also considered.
%

Figures \ref{fig:ber_users_all_snr20} and
\ref{fig:ber_users_all_snr5} present the simulation results for
$E_b/N_0=20 dB$ and $E_b/N_0=5 dB$, respectively.
For $E_b/N_0=20dB$, 
MUI is the dominant impairment. 
The performance of the ID and FG3 detectors gradually
deteriorate 
as the number of users is increased from 3 to 40. FG3 exhibits
better behavior throughout.
The LDPC coded system with CFG3 detector performs quite differently.
No errors are observed until the number of users reaches a certain
threshold, 33 in this simulation. Once the threshold is passed,
performance degradation is steep. Exact position of the threshold
and its behavior depend on system parameters.
%

%
In the second simulation, shown in Figure
\ref{fig:ber_users_all_snr5}, the additive noise is not negligible.
The CFG3 detector behaves much like before: as long as the number of
users is below a certain threshold, performance degradation is quite
moderate. Also notice that the threshold moved to the left due to
the existence of noticeable noise. The ID and FG3 detectors perform
much worse than the CFG3 detector with FG3 being only slightly
superior to the ID detector.

%
%
\subsection{Performance dependency on iterations}
%
%
%
%
The influence of the number of iterations on performance
has so far been overlooked; 
a fixed number of $8$ iterations was used throughout.
Notably, increasing the number of iterations benefits the LDPC-based
system more than the alternatives.

\begin{figure}[t]
    \centering
        \includegraphics[width=\columnwidth]
            {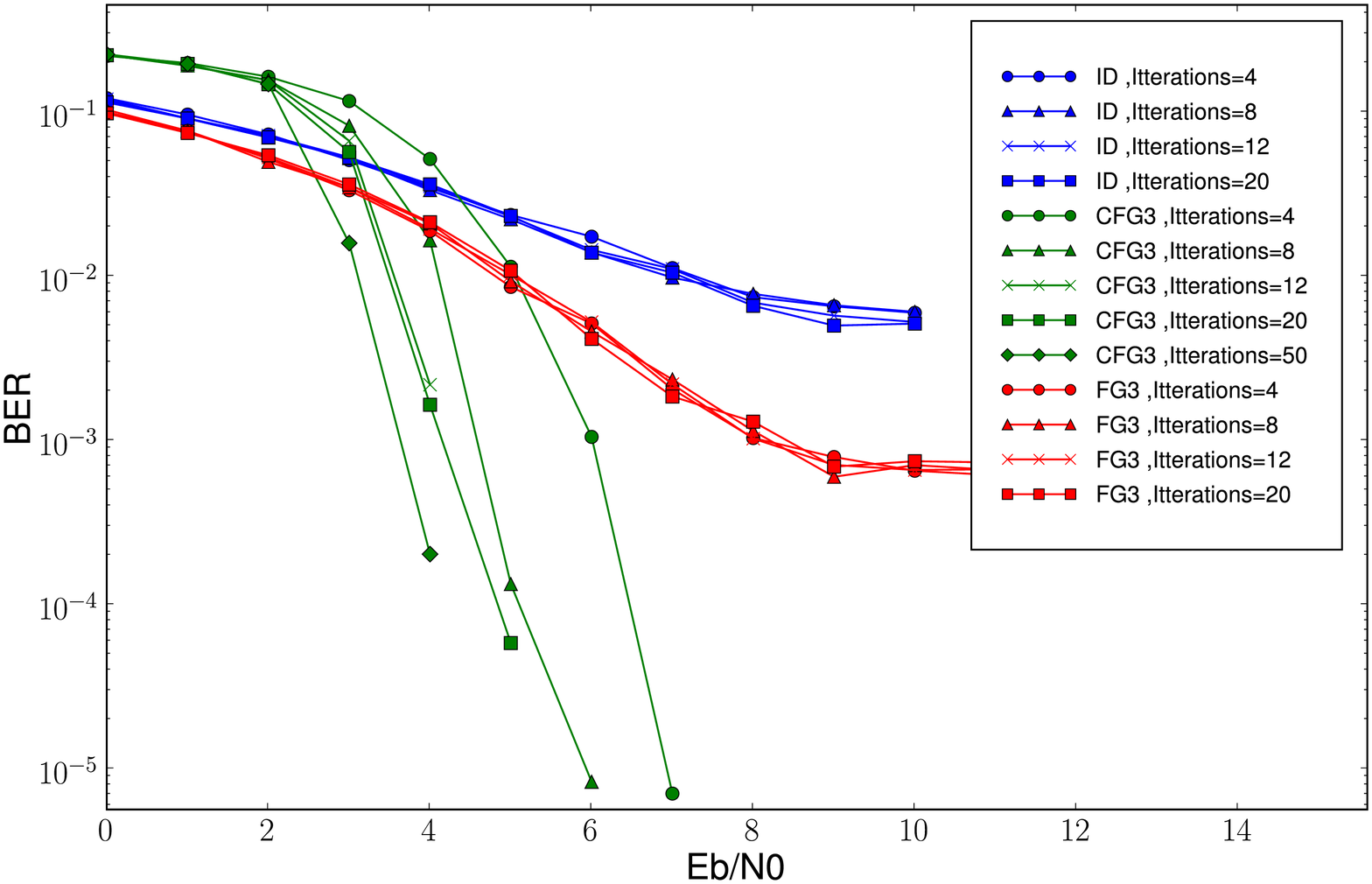}
    \vspace*{-0.8cm}
    \caption{Performance dependency on number of iterations, $K=20$ users}
    \label{fig:iter_20user}
\end{figure}

As an illustrative example Figure \ref{fig:iter_20user} presents the
BER performance of 
the three detectors, as a function of $E_b/N_0$, for $K=20$ users
and different number of iterations.
%
%
%
The performance of the ID and FG3 detectors converge in only 4
iterations.
More iterations are required for the LDPC-based system to converge.
In the single-user case (simulation results not shown), the number
of  iterations required for the LDPC-based system to converge is
smaller than in the multi-user case. One may argue that this follows
from the fact that the graph associated with the multi-user case is
much more involved.

%
%
\subsection{Some after simulation comments}
%
%
%

In a single-user system, the user can clearly employ all the time
slots for its own usage. 
In the case of high SNR, with two-level modulation, the channel
capacity is upper bounded by
$1[bit/chip]$. Since each frame provides $N_c$ transmit
opportunities, the capacity of the single-user system is upper
bounded by $C=N_c [bits/frame]$.
%
Let us now consider the CFG3-based system with $K$ users. The code
rate used  was $0.46$ and $N_c=20$. The CFG3 detector hardly
produced any bit errors
as long as the number of users satisfied $K<33$. The overall system
throughput in this case is  $C=14.72 [bits/frame]$ as compared to
the above upper bound of $C=N_c=20 [bits/frame]$.
With an optimized LDPC code of greater length, the obtained
throughput is  expected to grow closer to this bound.

\section{Conclusions}
\label{chap:conclusions}

%
%
%
This work is concerned with iterative receivers for a multiuser TH-IR system.
In particular, we study the gain achieved by introducing coding into the originally uncoded system.
Several iterative multi-user detectors have been presented based on
factor graph representation of the complete system. These detectors are
general and can support any binary linear coding scheme.
Yet another strength of the proposed
approach is that the graph used may be extended to account for multipath components (owing to the nature
of the UWB channel).
%
%
This is an interesting topic for future work.

Simulation results for the above mentioned detectors are presented
for several codes including LDPC codes. It is demonstrated that for
$20$ users, BER of $10^{-3}$ can be achieved with higher system rate
and more than $4dB$ gain in $E_b/N_0$ when using an LDPC-coded
system. Furthermore, it is shown that the achievable performance of
the original system quickly saturates at relatively poor BER.
%


%
%
%

\bibliography{iterative_uwb_ir}
\bibliographystyle{unsrt}

\end{document}